# Conceptual Design of In-Vessel Divertor Coils in DTT

Emilio Acampora[1], Raffaele Albanese[1,2,3], Roberto Ambrosino[1,2,3], Antonio Castaldo[4], Paolo Innocente[5], Vincenzo Paolo Loschiavo[6]

[1] Università degli studi di Napoli Federico II, via Claudio 21, I-80125, Napoli, Italy
[2] CREATE-ENEA, via Claudio 21, I-80125, Napoli, Italy
[3] DTT S.C. a r.l., Frascati, Italy
[4] ENEA, C.R. Frascati, Via Enrico Fermi 45, 00044 Frascati, Rome, Italy
[5] Consorzio RFX, Corso Stati Uniti 4, Padova, 35127, Italy
[6] Università degli studi del Sannio, Piazza Roma 21, 82100, Benevento, Italy

Corresponding Author: R. Albanese; raffaele.albanese@unina.it; CREATE - Univ. Napoli Federico II, via Claudio 21, I-80125, Napoli, Italy

*Abstract*— **The Divertor Tokamak Test (DTT) facility will be equipped with in-vessel divertor coils able to locally modify the flux surfaces in the divertor region. When the first DTT divertor had not been selected, four in-vessel divertor coils were considered, with 10 turns each, fed by independent 4-quadrant SCR (thyristor) 0.5 kV - 5 kA - power supplies. This configuration was able to meet the design criteria in terms of requested performance and operational constraints, but one of the coils is not compatible with the geometry of the first DTT divertor, and another coil, scarcely efficient, can be removed. This paper discusses the design criteria, illustrates the possible revised layout, and illustrates the performance that can be achieved.**

## I. INTRODUCTION

The Divertor Tokamak Test (DTT) facility [1] has been conceived as a flexible test bed capable to tackle plasma exhaust issues in a fairly integrated fashion, according to the EU Fusion Roadmap. DTT will firstly assess the performance of a conventional divertor for a single null (SN) plasma configuration with dimensionless parameters similar to ITER and DEMO. DTT will then explore and test the use of alternative materials and new divertor concepts.

Thus, one of the DTT requirements is the flexibility in the divertor region, so as to test alternative magnetic configurations, including X-Divertor (XD), Snowflake (SF), Negative Triangularity (NT), and "long leg" (LL) [2].

DTT will be equipped with in-vessel divertor coils able to locally modify the flux surfaces in the divertor region [1]. This paper illustrates the proposal for their conceptual design, describing the performance in terms of impact on the local magnetic field, strike-points sweeping, control of secondary nulls.

The reference proposal for these coils was worked out in 2021. However, the geometry of the first divertor in collaboration with EUROfusion [2] has introduced some modifications in the reference plasma scenarios and in the space available for the in-vessel coils. The conceptual design of the in-vessel divertor coils has then been reviewed as discussed in the present paper, which also reports the technical specifications for the power supplies for the divertor coils.

## II. DESIGN CRITERIA

### II.A Control of strike-points, SOL and secondary X-point

On the time scale of 250 ms, the divertor coils should be able to provide:

- strike point displacement normal to the leg in the single null (SN) configuration of at least 15 mm inboard and 24 mm outboard
- effective performance in XD configuration control: 120 mm movement of the secondary X-point and large displacement (60 mm) of outboard strike points

### II.B Assistance in X-point formation & diverted-limited transition in ramp-down

The magnetic field needed for DTT scenarios can be achieved using only the superconducting poloidal field coils. However, the in-vessel divertor coils could effectively be used for fine control of the diverted configurations at low plasma current and during fast transients.

### II.C Strike-point sweeping

The divertor coils should be used for strike-point sweeping in case of attached plasmas to get a 33% reduction of the peak flux density at a frequency of 4 Hz without significant changes in the plasma boundary.

### II.D Constraints

To guarantee coil and power supply protection, the design has to take into account suitable limits for current (5 kA), voltage (500 V), temperature (about 50 K increase), force (compatible with the admissible loads, about 7.5 MN each coil).

## III. LAYOUT

### III.A Previous layout (2021)

In the previous layout [4], when the first DTT divertor had not been selected, four in-vessel divertor coils were considered, with 10 turns each (Fig. 1): IVC1 close to the internal rail, IVC2 and IVC3 attached to the vacuum vessel close to the lower port, IVC4 in the outboard divertor region.

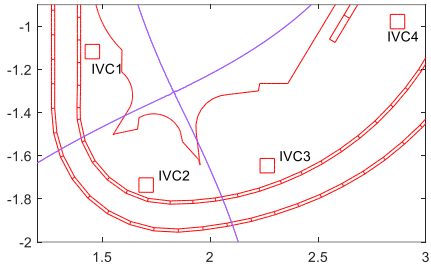

Fig. 1. Four in-vessel divertor coil (10 turns each) in the previous layout (2021), before the release of the first DTT divertor design.

### III.B Revised layout (2022)

There are stringent constraints for the location of the coils. They cannot be placed in front of the ports for compatibility with diagnostics, additional heating, and maintenance. In addition, they should be attached to the vessel shell for electromechanical reasons, otherwise they would be vulnerable to the disruptions.

Due to its location in the previous layout, shown in Fig. 1, the efficiency of IVC4 was poor, with potential electromechanical problems. Therefore, IVC4 has been removed in the revised layout.

The position of IVC1 shown in Fig. 1 was not compatible with the baseline divertor. In Option A (Fig. 2a), there is no interference, but the location of IVC1 under the inner divertor rail does not fit with the assembly procedure [3]. Option B (Fig. 2b) is then being considered, with 8 turns in IVC1a above the rail in series with 8 turns in IVC1b (Fig. 2b). In both options, the number of turns of IVC2 and IVC3 has been increased from 10 to 12.

In view of possible subsequent advanced divertor configurations, Option 2022B is more flexible, since IVC1a and IVC1b might also be independently fed. The possible subsequent divertors will hardly be larger than the DTT divertor presented in [3] and shown in Fig. 3. The in-vessel coil efficiency will be certainly better in case of lower positions of the X-point.

In case of future divertors optimized for plasmas with the legs shifted outwards, e.g., with a negative triangularity, the inboard coils would not be very effective, but an additional in-vessel coil might be added just above the oblique lower port.

### IV. PERFORMANCE
#### IV.A Disruptions

The fast electromagnetic transients due to plasma disruptions may induce large currents in the in-vessel components [5]. The most dangerous event expected for the DTT divertor coils is a downward disruption of a 5.5 MA plasma in 4 ms after a vertical displacement event (VDE). These may yield excess currents with large mechanical loads on the coils and possible damages to the equipment of the electric circuits. The insertion of additional resistors and inductors in series can limit the induced currents from 40 to 12 kA during disruptions (Fig. 3), yielding electromechanical loads compatible with the supports.

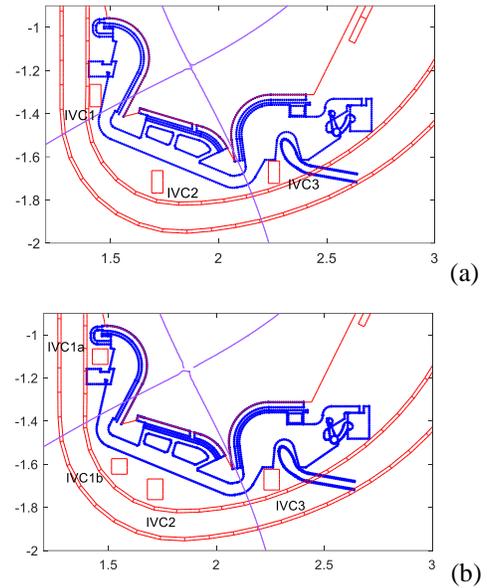

Fig. 2. Revised layout (2022): a) option A, with three in-vessel divertor coils (12 turns each); b) option B, with three independent circuits: IVC1a (8 turns) and IVC1b (8 turns) in series, IVC2 (12 turns), and IVC3 (12 turns).

#### IV.B Thermal load

The insulated copper conductor in a stainless-steel protection tube shown in Fig. 4 is being considered for the manufacture of the DTT in-vessel coils [6]. The final decision will be made taking into account the effect of the maximum temperature achieved during vacuum vessel baking.

Assuming a tolerable temperature increase $\Delta T=50K$ due to the ohmic losses in normal operation during the pulse, the tolerable value of $I^2_{rms} \cdot \Delta t$ would be 288 $kA^2$ s, corresponding for instance to an RMS current value $I_{rms}$=1.7 kA for a duration $\Delta t$=100s, or 2.4 kA for 50 s.

#### IV.C Strike point sweeping

The strike points are the intersections of the separatrix with the divertor plates. Strike point sweeping is a power exhaust strategy aimed at enlarging the divertor area affected by the plasma scrape-off layer (SOL), by imposing a periodic movement of the plasma divertor legs at a desired frequency without significant changes of the plasma boundary.

For the single null (SN) configuration, there is a fair decoupling between divertor legs and plasma boundary in the frequency range 1÷10 Hz. Using three 4-quadrant SCR (thyristor) power supplies, each with limit current I=5 kA and voltage V= 0.5 kV, a movement of ±22 mm at 5 Hz can be obtained for the outboard leg (±15 mm for the inboard leg) in Option 2022B. Considering the poloidal angles of inclination of the legs at the outer and inner target ($\alpha_{out}$=18 deg, $\alpha_{in}$=35 deg), the motion is larger along the targets, and one order of magnitude larger than the oscillation of the plasma boundary. With a SOL power of 25 MW and a flux decay length of 0.8 mm at the plasma boundary att the equatorial plane [2], the peak power density could be reduced from 40 MW/m$^2$ to less than 10 MW/m$^2$, with a periodic strike-point motion of $\pm 25$ mm inboard and $\pm 70$ mm outboard (Fig. 5).

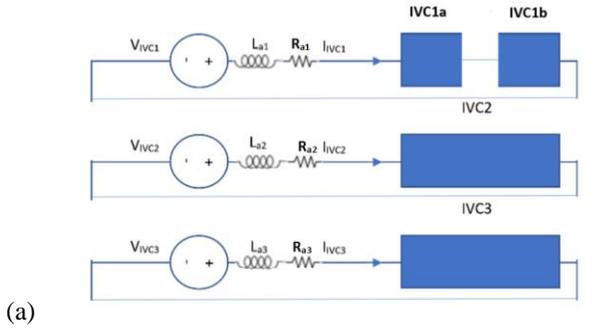

(a)

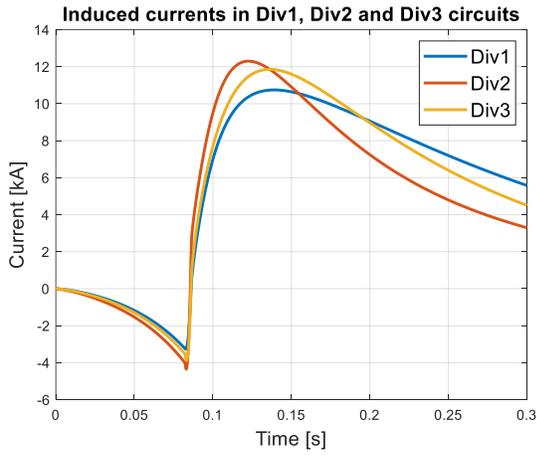

(b)

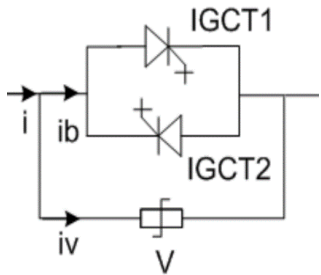

(c)

Fig. 3. In-vessel divertor circuit (option 2022B), with three independent SCR power supplies (5 kA, 0.5 kV each): a) circuit scheme with resistances and inductances in series; b) overcurrents due to a lower disruption of a 5.5 MA plasma in 4 ms following a VDE (here $R_{1a} = R_{2a} = R_{3a} = 0$, $L_{1a} = 2$ mH, $L_{2a} = 0$, $L_{3a} = 1$ mH); c) IGCT circuit breakers, which can be added in series to provide further protection.

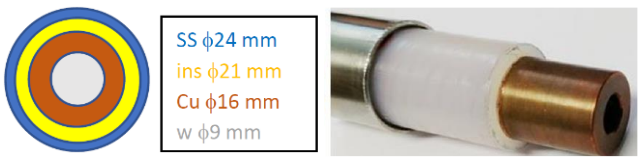

Figure 4. Insulated conductor in a steel protection tube similar to AUG [6]. The various layers are water, copper, insulator (perfluoroalkoxy copolymer resin), stainless-steel.

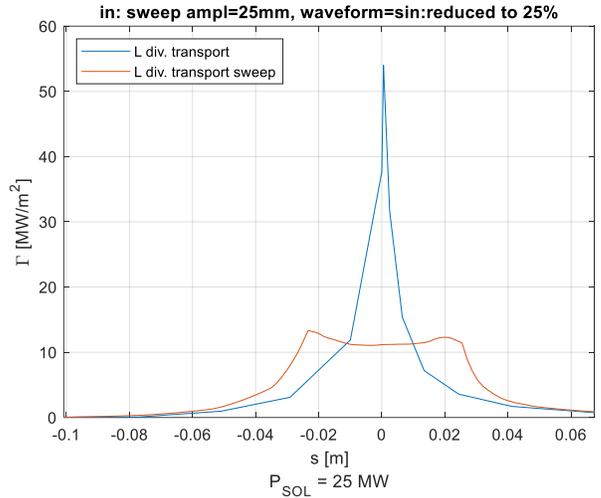

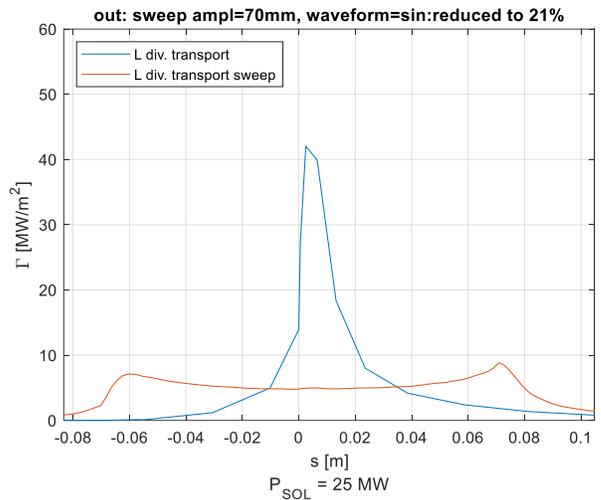

Figure 5. Peak flux density reduction with strike-point sweeping at 4 Hz (25 MW SOL power, 0.8 mm power decay length) for 2022 Option B layout.

### *IV.D Fine control of the divertor region*

This analysis was carried out for the 2021 layout. However, they are still significant, since the contribution of IVC4 was modest, and the relative position of targets and plasmas are very similar to the revised layout.

The strike point displacement normal to the leg for the single null (SN) is $\Delta s_{in} = 15$ mm, $\Delta s_{out} = 24$ mm. The displacement is enhanced by the factor of $1/\sin \alpha$ on the target: $\Delta s_{in}/\sin \alpha_{in} = 26$ mm, $\Delta s_{out}/\sin \alpha_{out} = 81$ mm (Fig. 6a).

For the X-Divertor (XD) configuration the secondary X-point can be moved closer to the active X-point reducing the distance from 400 to 260 mm (Fig. 6b).

The control efficiency is obviously increased during ramp-down, ramp-up and X-null formation, due to the reduced value of the plasma current.

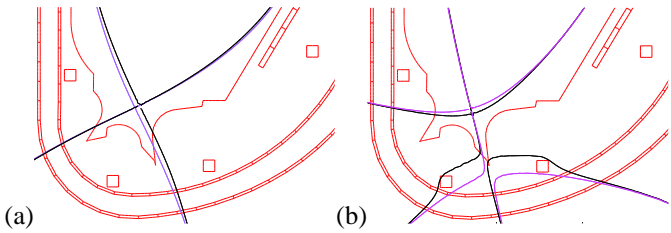

(a)    (b)

Figure 6. Displacement of the SN strike-points and XD secondary X-point using the in-vessel divertor coils of the 2021 layout: a) displacement of the SN outer strike point; b) displacement of the XD secondary X-point (the null point outside the plasma boundary) from 40 to 26 cm.

## V. Conclusions

In the previous layout, when the first DTT divertor had not been selected, four in-vessel divertor coils were considered. The position of IVC1, the innermost coil, was not compatible with the baseline divertor, whereas the efficiency of IVC4, the outermost coil, was poor. For these reasons, a revised layout (Option 2022B) is examined, in which IVC1 is split in two coils in series, the number of turns has been increased (8+8 in IVC1, 12 in IVC2, 12 in IVC3). The coils are fed by three independent 4-quadrant SCR (thyristor) 0.5 kV - 5 kA power supplies. Circuit protections are considered to limit the overcurrents during disruptions. A thorough analysis shows that this layout was able to meet the specifications in terms of performance and constraints requested by the design criteria.

## VI. Acknowledgments

This work has been carried out within the framework of the EUROfusion Consortium, funded by the European Union via the Euratom Research and Training Programme (Grant Agreement No 101052200 — EUROfusion). Views and opinions expressed are however those of the author(s) only and do not necessarily reflect those of the European Union or the European Commission. Neither the European Union nor the European Commission can be held responsible for them.